\documentstyle[twocolumn,aps,psfig]{revtex}
\begin{document}             
\draft
\wideabs{          
\title{Giant enhancement of the thermal Hall conductivity $\kappa_{xy}$ in the 
superconductor $\rm YBa_2Cu_3O_7$} 
\author{Y. Zhang$^1$, N. P. Ong$^1$, P. W. Anderson$^1$, D. A. Bonn$^2$, R. Liang$^2$ 
and W. N. Hardy$^2$}      
\address{$^1$Joseph Henry Laboratories of Physics, Princeton University, Princeton, 
New Jersey 08544}
\address{$^2$Department of Physics, University of British Columbia, Vancouver, 
Canada.}
\date{\today}      


\maketitle                   

\begin{abstract}
In high-purity crystals of $\rm YBa_2Cu_3O_7$, the quasiparticle (qp) lifetime $\tau$ 
and the (weak-field) thermal Hall conductivity $\kappa_{xy}$ undergo dramatic 
increases below 90 K. We present a detailed picture of the behavior of $\kappa_{xy}$ 
at low temperature, in particular its scaling properties, which are directly relevant 
to the issue of whether Landau quantization of the qp states occurs.
\end{abstract}
}				
The problem of excitations of the superconducting condensate in the cuprates at low 
temperatures is of strong current interest.  In a $d$-wave superconductor, the energy-
momentum dispersion of quasiparticles near a node is Dirac-like.  The effect of an 
intense magnetic field on the quasiparticle (qp) states is an interesting open 
question \cite{Gorkov,Anderson,Janko,Kopnin,Melnikov,Franz3}.  Landau quantization of 
the qp states, first proposed by Schrieffer and Gorkov \cite{Gorkov}, has been 
recently re-derived using different arguments \cite{Anderson,Janko,Kopnin}.  However, 
the case against Landau-level formation has also been argued \cite{Melnikov,Franz3}.  

A second problem is the temperature dependence of the qp mean-free-path $\ell$ (in 
zero field) close to $T_c$.  Transport evidence from thermal conductivity \cite{Yu}, 
microwave and teraHertz experiments \cite{Bonn,Nuss,Orenstein}, and thermal Hall 
conductivity \cite{Krish95,Zeini,Krish99} point to a sharp increase in the qp lifetime 
just below $T_c$.  Recent high-resolution angle-resolved photoemission (ARPES) 
experiments \cite{Valla,Kaminski} have started to address the lifetime issue as well, 
but with conflicting results (see below).  

These issues reflect the strong interest in the low-lying excitations of the $d$-wave 
superconductor.  While microwave absorption and ARPES experiments provide valuable 
information on the quasiparticles, they are less effective in a field.  For in-field 
experiments, teraHertz techniques \cite{Orenstein} and the thermal Hall effect 
\cite{Krish95,Zeini,Krish99}, in particular, have emerged as powerful probes of qp 
transport.  In a field, the qp heat current develops a transverse component that is 
observed as a thermal Hall conductivity $\kappa_{xy}$ (by contrast, phonons do not 
display a Hall effect since they are charge-neutral).  Hence, $\kappa_{xy}$ {\em 
selectively} senses the qp current alone \cite{Krish95}.  To fully exploit this 
technique at low temperatures, however, samples with a very long $\ell$ are needed.

A recent innovation is the growth, using $\rm BaZrCO_3$ (BZO) crucibles, of crystals 
of $\rm YBa_2Cu_3O_y$ (YBCO) with nearly perfect crystalline order (from x-ray rocking 
curves \cite{Liang}) and very low impurity concentration.  The step-wise improvement 
in crystal quality results in strong enhancements of the qp lifetime $\tau$.  A number 
of novel features of qp heat transport become apparent in these crystals.  The weak-
field $\kappa_{xy}$ undergoes a remarkable thousand-fold increase between $T_c$ and 30 
K.  Below 30 K, the curves of $\kappa_{xy}$ vs $H$ provide new, specific information 
on scaling behavior at low $T$ \cite{Simon}.  Both features are directly relevant to 
the two issues mentioned above.  

\begin{figure}[h]
\centerline{\psfig{figure=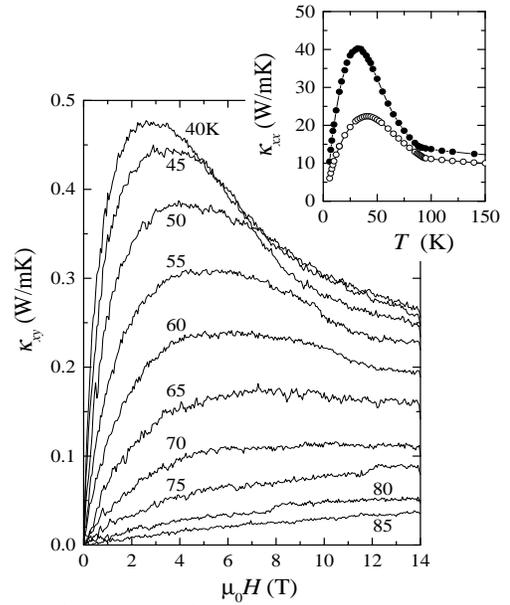,height=3.2in,width=2.7in}}
\caption{(Main Panel) The thermal Hall conductivity $\kappa_{xy}$ vs. $H$ in BZO-grown  
$\rm YBa_2Cu_3O_{6.99}$ ($T_c = 89$  K) at temperature from 85 to 40 K.  As $T$ 
decreases below $T_c$, the initial slope $\kappa^0_{xy}/B$ increases sharply.  The 
prominent peak in $\kappa_{xy}$ below 55 K is a new feature in BZO-grown YBCO.  The 
inset compares the zero-field $\kappa_{xx}\equiv \kappa_a$ in the BZO-grown crystal 
(solid circles) with a detwinned non-BZO grown crystal (open). 
}
\label{khigh}
\end{figure}

\begin{figure}[h]
\centerline{\psfig{figure=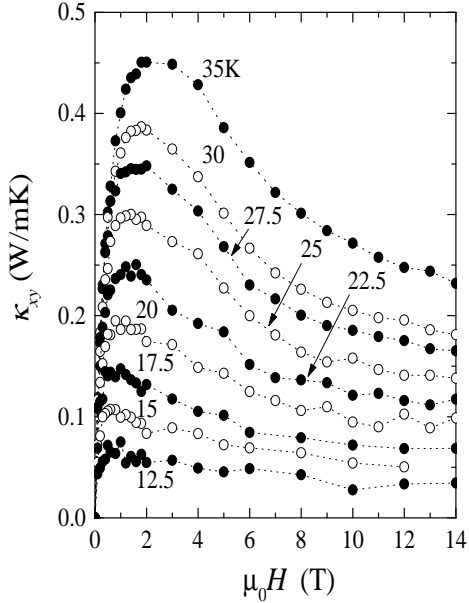,height=3.2in,width=2.7in}}
\caption{The thermal Hall conductivity $\kappa_{xy}$ vs. $H$ in BZO-grown $\rm 
YBa_2Cu_3O_{6.99}$ between 35 and 12.5 K.  Below 28 K, the peak value varies as $T^2$ 
(see text).  
}
\label{klow}
\end{figure}

Among the cuprates, 90-K $\rm YBa_2Cu_3O_7$ displays the largest in-plane thermal 
conductivity anomaly.  In BZO-grown crystals, this anomaly is further enhanced, as 
shown in the inset in Fig. \ref{khigh}.  The longitudinal thermal conductivity 
$\kappa_{xx}$ ($-\nabla T\parallel {\bf a}$) in BZO crystals (solid circles) attains a 
peak value that is $\sim 80 \%$ larger than that seen in typical, non-BZO detwinned 
crystals (open circles).  To isolate the qp current, we turn to $\kappa_{xy}$.
The main panel of Fig. \ref{khigh} displays traces of $\kappa_{xy}$ vs. field $B$ from 
85 to 40 K \cite{expt}.  As in earlier studies \cite{Krish95,Krish99}, the initial 
slope $\kappa^0_{xy}/B\equiv \lim_{B\rightarrow 0} \kappa_{xy}/B$ increases very 
rapidly as the temperature $T$ falls below $T_c$. Further, the curves are strongly 
non-linear in $H$.  Both features reflect a $\tau$ that increases rapidly with 
decreasing $T$. An important new feature, absent in previous studies, is the prominent 
`overshoot' that produces a maximum in $\kappa_{xy}$ at the field scale $H_{max}$.  As 
$T$ falls below 40 K (see Fig. \ref{klow}), the peak continues to narrow.  For later 
reference, we note that, over a broad range of temperatures (10 $<T<$70 K), $H_{max}$ 
varies as $T^2$.  Moreover, at low temperatures ($T<$28 K), the peak magnitude 
$\kappa^{max}_{xy}$ also scales as $T^2$. 

\begin{figure}[htb]
\centerline{\psfig{figure=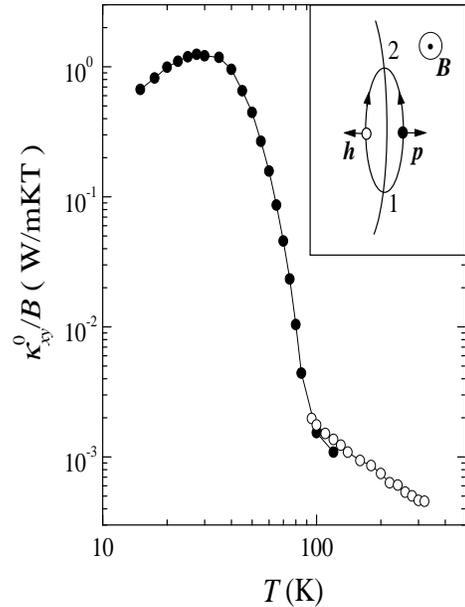,height=3.2in,width=2.7in}}
\caption{The $T$ dependence of the initial Hall slope $\kappa^0_{xy}/B$ in BZO-grown 
YBCO (solid circles).  Between $T_c$ and 30 K, $\kappa^0_{xy}/B$ increases by $10^3$.  
The $1/T$ dependence of $\kappa_{xy}^0/B$ above $T_c$ (measured in a non-BZO grown 
YCBO) is shown as open circles. The inset shows a qp energy contour on the Dirac cone.  
Group velocities on the particle- ($p$) and hole-like ($h$) branches are indicated.}
\label{kxyT}
\end{figure}

The initial slope $\kappa^0_{xy}/B$, plotted as solid circles in Fig. \ref{kxyT}, 
undergoes a thousand-fold increase between $T_c$ and 30 K (the $T$-linear variation of 
$\kappa_{xy}$ above $T_c$ are displayed as open circles \cite{Zhang1}).  We now show 
that this giant enhancement is driven by a 100-fold increase in the qp lifetime.  

To extract the zero-field mean-free-path (mfp) $\ell$ from $\kappa^0_{xy}/B$, we apply 
the Boltzmann-equation approach \cite{BRT}, which should be valid in the {\em weak}-
field regime $\omega_c\tau\ll 1$ ($\omega_c$ is the cyclotron frequency). In terms of 
the `qp heat capacity' $c_e = T^{-1}\sum_{\bf k}(-\partial f/\partial E_{\bf k})E_{\bf 
k}^2$ where $E_{\bf k}$ is the qp energy, the zero-$H$ thermal conductivity may be 
written as $\kappa_e = c_e \langle v\ell\rangle/2$, with the group velocity ${\bf v_k} 
= \nabla E_{\bf k}/\hbar$.  (Close to a node ${\bf k}^*$, the qp energy may be 
approximated as $E_{\bf q} = \hbar\sqrt{(v_fq_1)^2 + (v_{\Delta}q_2)^2}$, where $v_f$ 
and $v_{\Delta}$ are velocity parameters normal and parallel to the FS, and ${\bf q = 
k - k}^*$.)  

The thermal Hall conductivity is related to $\kappa_e$ by $\kappa_{xy} = 
\kappa_e\tan\theta$.  We assume that, in the weak-field limit, the thermal Hall angle 
$\tan\theta$ is proportional to $\omega_c\tau$, viz. 
\begin{equation}
\tan\theta = \eta\omega_c\tau = \eta\ell/k_F\ell_B^2,\;\;(B\rightarrow 0)
\label{tan}
\end{equation}
where the magnetic length $\ell_B = \sqrt{\hbar/eB}$. The parameter $\eta$ is less 
than 1 if $\ell$ is anisotropic around the FS.  To obtain $\tan\theta$ \cite{Krish99}, 
we first fit the profile of $\kappa_{xx}$ vs. $H$ to the empirical expression 
$\kappa_{xx}(B,T) = \kappa^0_e(T)/[1+p|B|^\mu] + \kappa_{bg}(T)$, where the background 
term $\kappa_{bg}(T)$ is $H$-independent and identified with the phonon contribution.  
The initial Hall angle is then obtained as \cite{Krish99}
$\tan\theta = \lim_{B\rightarrow 0}\kappa_{xy}(B)/[\kappa_{xx}(B)-\kappa_{bg}].$  This 
procedure allows us to extract $\tan\theta$ (hence, $\ell$ using Eq. \ref{tan}).  

As a consistency check, we adopt a second way to obtain $\ell$ from $\kappa^0_{xy}$ 
that relies on measurements of the electronic heat capacity $c_e$.  Using Eq. 
\ref{tan}, we may write 
\begin{equation}
\kappa^0_{xy} = \frac{c_e v_f\ell^2\eta}{4k_f\ell_B^2}
\label{kxy}
\end{equation}

In a $d$-wave superconductor, $c_e = \alpha_c T^2$ for $T<T_c$.  Using the measured 
value  $\alpha_c \simeq 0.064\; {\rm mJK^{-3}mol^{-1}}$ \cite{Phillips}, we may invert 
Eq. \ref{kxy} to find $\ell$.  We find that the values of $\ell$ obtained from the two 
methods share the {\em same} $T$ dependence, but differ by a fixed factor of 1.5 if 
$\eta=1$.  By adjusting $\eta$ to 0.6, we obtain numerical agreement between the two 
methods. 

\begin{figure}[htb]
\centerline{\psfig{figure=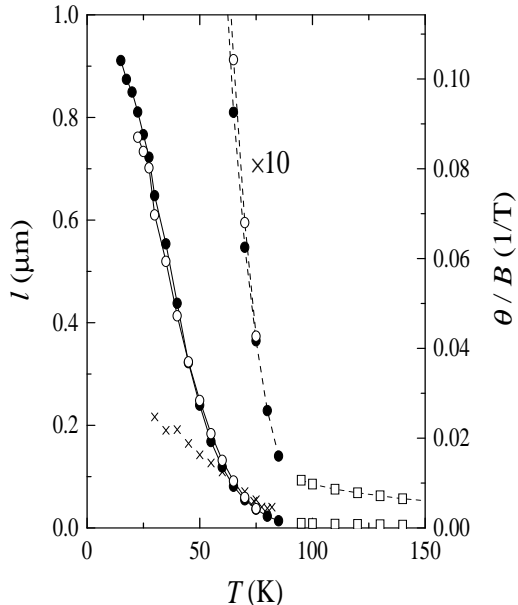,height=3.2in,width=2.7in}}
\caption{The zero-field mean-free-path $\ell$ extracted from the weak-field Hall angle 
$\tan\theta$ (open circles), and from Eq. \ref{kxy} (closed).  The {\em equivalent} 
values of $\theta/B$ are shown on the right scale.  The symbols ($\times$) represent 
$\tan\theta$ measured in a non-BZO detwinned YBCO crystal (Krishana {\em et al.} 
[14]).  The expanded scale (dashed lines) highlights the steep increase below $T_c$.  
To extract $\ell$, we used the values $\eta=0.60$, $v_f = 1.78\times 10^7$ cm/s, and 
$k_f = 0.8 A^{-1}$. }
\label{ell}
\end{figure}

Figure \ref{ell} shows the $T$ dependence of $\ell$ derived from the two methods.  The 
agreement between the two sets of data is evidence that our assumption Eq. \ref{tan} 
is physically reasonable.  Remarkably, between $T_c$ and 20 K, the mfp increases by a 
factor of $\sim$120 from 80 A to 1 micron.  In the expanded scale, we show that this 
increase is abrupt, starting slightly below $T_c$.  

[For comparison, $\tan\theta$ measured previously in a non-BZO crystal \cite{Krish99} 
is shown as $\times$.  Based on the higher sensitivity and broader range in $T$ in the 
present experiment, we now conclude that $\tan\theta$ does {\em not} lie on the 
extrapolated curve for the electrical Hall angle $\tan\theta_e$.]

Beyond the weak-field regime, we need a fully microscopic description of the qp 
thermal Hall current to properly analyze $\kappa_{xy}$ vs. $H$.  As the theoretical 
situation is unsettled, we adopt instead scaling arguments \cite{Simon}.  This 
approach reveals some rather striking features in the data.  

For states close to the node ${\bf k}^*$, the linear energy dispersion $E= \hbar 
\bar{v} q$ ($\bar{v}$ is an average velocity) implies a general relation between 
$k_BT$ and the magnetic length $\ell_B$ at a characteristic field scale $B_s$, viz. 
\begin{equation}
k_BT = \hbar\bar{v} \sqrt{\frac{eB_s}{\hbar}}.
\label{vbar}
\end{equation}

In addition to this general relation, Simon and Lee \cite{Simon} have proposed that, 
at low $T$ ($<$30 K for YBCO), the magnitude of $\kappa_{xy}$ should scale as
\begin{equation}
\kappa_{xy}(H,T) \sim T^2F_{xy}(\sqrt{H}/\alpha T),
\label{Fxy}
\end{equation}
where $\alpha \equiv k_B/\bar{v}\sqrt{e\hbar}$, and $F_{xy}(u)$ is a scaling function 
of the dimensionless parameter $u = \sqrt{H}/\alpha T$.  Hence, plots of 
$\kappa_{xy}/T^2$ versus $\sqrt{H}/T$ should collapse to the universal curve 
$F_{xy}(u)$.

\begin{figure}[htb]
\centerline{\psfig{figure=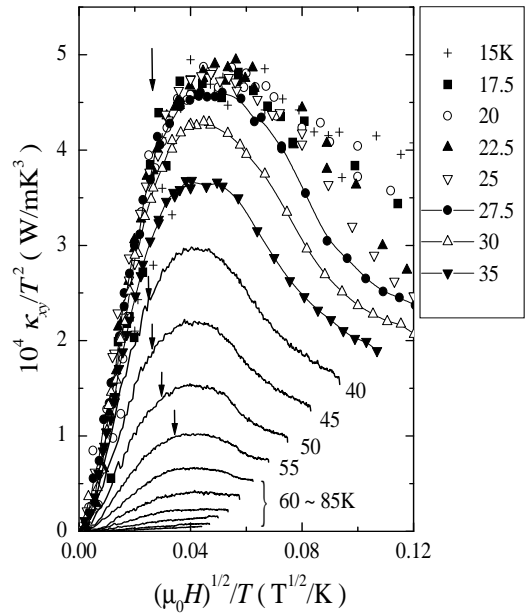,height=3.2in,width=2.7in}}
\caption{Simon-Lee scaling plot of $\kappa_{xy}/T^2$ versus $\sqrt{H}/T$ (Eq. 
\ref{Fxy}).  Below 28 K, the curves collapse onto a `universal' curve $F_{xy}(u)$.  
Above 28 K, scaling is violated.  However, the peaks still occur at the same $x$-
coordinate ($\sqrt{H_{max}}/T = 0.042$).  The arrows indicate the field scale 
$H_{arc}$.
}
\label{scale}
\end{figure}

We proceed to plot our results in this way in Fig. \ref{sqrt}.  While the curves above 
28 K are spread out, the ones below collapse onto a common curve for $H< H_{max}$.  
The data taken at 25 K (and below) collectively determine the form of $F_{xy}(u)$.  
Its most notable feature is the nominally straight segment that extends from $u\simeq 
0$ to just below $u_0\equiv \sqrt{H_{max}}/\alpha T$, i.e. $F_{xy}(u)\sim u$ for 
$0<u<u_0$.  This simple form for $F_{xy}$ implies that, below 25 K and for 
$H<H_{max}$, $\kappa_{xy}$ reduces to the form
\begin{equation}
\kappa_{xy}(H,T) = C_0T\sqrt{H},
\label{sqrt}
\end{equation}
where the constant $C_0 = 1.51\times 10^{-2}$ in SI units.  Remarkably, when Eq. 
\ref{sqrt} applies, the magnitude of $\kappa_{xy}$ is just proportional to 
$T\sqrt{H}$, and is insensitive to all transport quantities such as $\ell$ and 
$\theta$.  This interesting result has not been anticipated theoretically.

At larger values of $u$, $F_{xy}$ attains a maximum value $F^0_{xy}$ before falling 
slowly.  The $T^2$ dependence of the peak value $\kappa^{max}_{xy}$ noted earlier in 
Fig. \ref{klow}, is now seen to be a simple consequence of scaling behavior (i.e. 
$\kappa^{max}_{xy}\sim T^2F^0_{xy}$).

Above 28 K, Simon-Lee scaling no longer holds.  Three field regimes are now apparent.  
In weak fields ($0<H<H_x$), $\kappa_{xy}$ is strictly linear in $H$.  Above  $H_x$, we 
enter a regime reminiscent of the $\sqrt{H}$ behavior at low-$T$ (the $H$-linear 
regime is too small to resolve below 28 K).  This intermediate regime appears as 
straight-line segments in Fig. \ref{scale}.  Finally, closer to $H_{max}$, 
$\kappa_{xy}$ deviates from $\sqrt H$ behavior, and goes through a broad maximum.  
Surprisingly, as noted earlier, the weaker scaling relation in Eq. \ref{vbar} 
continues to hold: Between 15 and 70 K, the maximum in $\kappa_{xy}$ occurs at the 
{\em same} $x$-coordinate in Fig. \ref{scale}, i.e.  $\sqrt{H_{max}}= 0.042\; T$.  
Substituting $H_{max}$ for $B_s$ in Eq. \ref{vbar}, we find that $\bar{v}\sim 8.0 
\times 10^6$ cm/s, which is close to the geometric-mean velocity 
$\sqrt{v_fv_{\Delta}}\sim 6.8 \times 10^6$ cm/s (with $v_f = 1.78\times 10^{7}$cm/s 
\cite{Valla} and $v_f/v_{\Delta}\sim 7$).    

We may estimate semiclassically the time that an excitation, in a field, takes to move 
from 1 to 2 along the arc (inset, Fig. \ref{kxyT}) by $\Delta t = (\hbar/eH) {\int^2_1 
ds_{\bf k}\mid{\bf v_k}\mid^{-1} }$, where $s_{\bf k}$ is the arc-length.  For this 
time to equal $\tau$, the field required is $H_{arc} = \pi E/(ev_{\Delta}v_f\tau)$.  
Using the measured $\ell\simeq v_f\tau$ at each $T$ and setting $E=k_BT$, we indicate 
$H_{arc}$ as arrows in Fig. \ref{scale}.  This rough estimate shows that the peak is 
related to the maximum arc length of the dominant energy contour on the Dirac cone.  
Hence, a detailed analysis of the Hall results should shed important light on the 
current debate about how vortices affect the qp spectrum.  The presence of Landau 
Levels \cite{Gorkov,Anderson,Janko,Kopnin} or absence \cite{Melnikov,Franz3} will 
presumably have a large effect on $\kappa_{xy}$.  Moreover, the direct measurement of 
the scaling function $F_{xy}$ (Fig. \ref{scale}) together with the other scaling 
features uncovered should stringently narrow the range of possibilities in this 
interesting problem.

The new results on $\kappa_{xy}$ also bear on the issue of the change in qp lifetime 
at $T_c$.  As discussed, $\ell$ derived from transport undergoes a steep increase just 
below $T_c$ \cite{Yu,Bonn,Nuss,Krish95}.  Recently, ARPES has attained enough 
resolution to probe the qp spectral peak along the nodal direction in $\rm 
Bi_2Sr_2CaCu_2O_8$.  Valla {\it et al.} \cite{Valla} find that the width $\Delta k$ 
($\sim 1/\ell_{ARPES}$) retains its $T$-linear dependence across $T_c$ (near $T_c$ 
$\ell_{ARPES}\simeq 25-30 {\rm \AA}$).  This appears to be in striking contrast with 
the transport results.  However, Kaminski {\it et al.} \cite{Kaminski} resolve a new 
feature of the qp peak that suggests a rather different picture.  Below $T_c$, the qp 
width becomes dramatically narrower {\em provided it is probed within 60 mV of the 
Fermi energy}.  They infer that well-defined qp states at the nodes exist only below 
$T_c$.  
The steep increase in $\ell$ shown in Fig. \ref{ell} is in close agreement with 
Kaminsky {\it et al.}  The data in Fig. \ref{ell} show that $\ell$ increases to 
$\simeq 1 \mu$m below 20 K (implying a peak 200 times narrower than the peaks resolved 
in the current ARPES studies). Hence, in high-purity YBCO, there are exceedingly sharp 
qp peaks in the spectral function that remain to be resolved and investigated. 
Understanding the abrupt appearance of the qp state below $T_c$, as implied by the 
steep increase in $\ell$ and $\kappa^0_{xy}/B$ near $T_c$, seems a key problem in the 
cuprates.

The research is supported by the U.S. National Science Foundation (Grant NSF-DMR 
9809483, at Princeton), and the Natural Science and Engineering Research Council 
(Canadian NSERC) and the Canadian Institute for Advanced research (CIAR) at U. British 
Columbia. N.P.O. also acknowledges support from the U.S. Office of Naval Research 
(Contract N00014-98-10081) and the New Energy and Industrial Tech. Develop. Org., 
Japan (NEDO). We thank T.V. Ramakrishnan for valuable comments.

%
%

\end{document}